\documentclass[journal,twocolumn]{IEEEtran}

\usepackage{cite}      

\usepackage{graphicx}  

\usepackage{amsmath}
\usepackage{amssymb}   
\begin{document}

\title{Absolute Calibration
of Analog Detectors by using Parametric Down Conversion}



%
\author{\authorblockN{Giorgio Brida\authorrefmark{1},
Maria Chekhova\authorrefmark{2}, Marco Genovese\authorrefmark{1},
Alexander Penin\authorrefmark{2},\\ Maria Luisa
Rastello\authorrefmark{1}and Ivano
Ruo-Berchera\authorrefmark{1}}\\
\authorblockA{\authorrefmark{1}Istituto Nazionale di Ricerca
Metrologica\\
Strada delle Cacce 91, 10135 Torino, Italy.
Email: ruo@ien.it}\\
\authorblockA{\authorrefmark{2}Physics
Department, M.V. Lomonosov State University, 119992 Moscow,
Russia. }
\thanks{The Torino group acknowledges  the contracts FIRB RBAU01L5AZ-002 and
PRIN 2005023443, the Fondazione San Paolo and the Regione Piemonte
(E14). The Russian group acknowledges the support of the Russian
Foundation for Basic Research (grant \# 05-02-1639) and the State
Contract of Russian Federation 2006-RI-19.0/001/595.}}


\maketitle

\begin{abstract}
\textbf{In this paper we report our systematic study of a
promising absolute calibration technique of analog
photo-detectors, based on the properties of parametric down
conversion. Our formal results and a preliminary uncertainty
analysis show that the proposed method can be effectively
developed with interesting applications to metrology.}
\end{abstract}

\begin{keywords}
\textbf{Detectors, calibration, optical parametric amplifier,
optical signal detection, optics, quantum theory}
\end{keywords}


%
 \IEEEpeerreviewmaketitle

\section{Introduction}
In optical radiometry primary standards are based on absolute
sources or detectors \cite{anal}. Synchrotron and blackbody
radiators are absolute sources. The relative uncertainty of both
these sources is about 1 part in $10^{3}$. Among the absolute
detectors, there exist the following two types: electrical
substitution radiometers (ESR) based on thermal effects, and
semiconductor photon detectors. Uncertainties of a few parts in
$10^{4}$ appear to be the limit of these techniques. However, such
a high accuracy of traditional calibration methods is reached only
for an appropriate choice of wavelength and special types of
detectors.

Recently the use of photons produced by means of spontaneous
parametric down-conversion (SPDC), where photons are emitted in
pairs strongly correlated in direction, wavelength and
polarization, has been exploited for the absolute calibration of
detectors in photon-counting mode
\cite{Burnham,klysh,malygin,alan,Brida1,Brida2}. This absolute
technique is becoming attractive for national metrology institutes
to realize absolute radiometric standards. In fact it relies
simply on the counting of events, involves a remarkably small
number of measured quantities, and does not require any reference
standards.

Because of the success of the SPDC scheme for calibrating
single-photon detectors, it is important to analyze the
possibility to extend this technique to higher photon fluxes for
calibrating analog detectors. A seminal attempt in this sense was
made in \cite{SP} following the theoretical proposal of
\cite{klysh}. Nevertheless, these results were limited to the case
of very low photon flux (as we will show later).

In this paper we report our systematic analysis of the measurement
method for increasing photon flux produced by means of SPDC,
showing how to estimate the quantum efficiency in a real analog
regime. Furthermore, we point out the intrinsic limitation of SPDC
for calibration when one moves towards fluxes requiring large
parametric gain. Thus, the stimulated parametric down conversion
(PDC) is considered as an alternative bright source of correlated
beams and we demonstrate that, under some opportune conditions,
the quantum efficiency can be estimated in this new regime.

\section{The absolute calibration technique}

The scheme for calibrating photon detectors by using parametric
down conversion is schematically depicted in Fig. 1. It is based
on the specific properties of this process, where a photon of the
pump beam (usually a laser beam) "decays" inside a non-linear
crystal into two lower-frequency photons, 1 and 2 , such that
energy and momentum are conserved ($\omega_{pump} = \omega_{1} +
\omega_{2}$, $\vec{k}_{pump} = \vec{k}_{1} + \vec{k}_{2}$).
Moreover, the two photons are emitted within a coherence time
$\tau_{coh}$ of tens of femtoseconds from each other. The process
can be spontaneous (SPDC) when no modes of radiation except the
pump modes are injected through the input face of the crystal. If
a seed mode $\vec{k}_{2}$ is injected, its presence stimulates the
process and many more photons of the pump are converted.

\begin{figure}[htbp]
\par
\begin{center}
\includegraphics[angle=0, width=9 cm, height=7 cm]{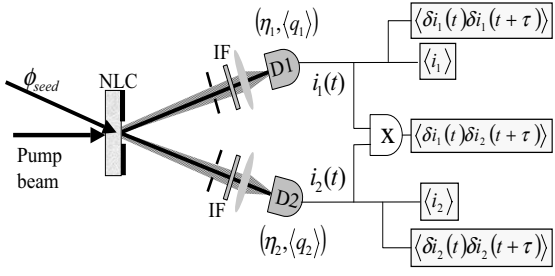}
\end{center}
\newpage\caption{Scheme for calibration of light detectors. The processing
of the detectors output indicated here is relative to the analog
regime.}
\end{figure}

In essence, the calibration procedure consists of placing a couple
of detectors $D1$ and $D2$ down-stream from the nonlinear crystal,
along the directions of propagation of correlated photon pairs for
a selected pair of frequencies. Since the incident photon fluxes
$F_{1}(t)$ and $F_{2}(t)$ are correlated within $10^{-13}$ s, the
fluctuations of the registered currents $i_{1}(t)$ and $i_{2}(t)$
are suppose to be strictly correlated. The non ideal quantum
efficiency of the detectors makes some photons missed sometimes by
$D1$ sometimes by $D2$, spoiling the correlation. The techniques
for estimating the quantum efficiency, both in counting and in
analog regime, consist in measuring this effect.


In the following the photodetection process in the analog regime
will be modelled as a random pulse train \cite{Soda}

\begin{eqnarray}
  i(t)=\sum_{n}q_{n}f(t-t_{n})\nonumber,
\end{eqnarray}
i.e. a very large number of discrete events at random times of
occurrence $t_{n}$. The pulse shape $f(t)$ is determined by the
transit time of charge carriers. We assume that $f(t)$ is a fixed
function with the characteristic width $\tau_{p}$ and a unit area
($f(t)$ has dimension $\textrm{s}^{-1}$). $\tau_{p}\sim 10$ ns
represents a typical value in analog detection. The pulse
amplitude $q_{n}$ is a random variable representing the electron
charge produced in the $n$-th event, in order to account for a
possible current gain by avalanche multiplication. The statistical
nature of the multiplication process gives an additional
contribution to the current fluctuations \cite{MNUAD}. In an ideal
instantaneous photocell response, without avalanche gain, all
values $q_{n}$ are equal to the charge $e$ of a single electron
and $f(t)\sim \delta(t)$. In the case of ideal quantum efficiency,
since the probability density of photo-detection at time $t$ at
detector $Dk$ ($k=1,2$) is related to the quantum mean value
$\langle F_{k}(t)\rangle$ of the photon flux (photons per second),
we calculate the average current output of $Dk$ as (see for
example \cite{QO})
\begin{eqnarray}\label{curr}
\langle i_{k}\rangle =\sum_{n}\langle q_{kn}f(t-t_{kn})\rangle
=\int dt_{k}\langle q_{k}\rangle f(t-t_{k})\langle
F_{k}(t_{k})\rangle\nonumber\\
\end{eqnarray}
where the factor $\langle q_{k}\rangle$ is the average charge
produced in a detection event. We have assumed the response
function for the two detectors to be the same,
$f_{1}(t)=f_{2}(t)=f(t)$.

Now we introduce the quantum efficiency $\eta_{k}$ of detector
$Dk$, defined as the number of pulses generated per incident
photon. In \cite{systematic} a real detector is modelled with an
ideal one ($\eta=1$) preceded by a beam splitter with transmission
coefficient equal to the quantum efficiency of the real detector
\cite{photon-noise}. Following the results reported there we can
perform in Eq. (\ref{curr}) the substitution $
F_{k}(t)\longrightarrow \eta_{k}F_{k}(t)$. Thus, being $\langle
F_{k}(t)\rangle$ time independent, we obtain:
\begin{equation}\label{curr-time}
  \langle i_{k}\rangle=\eta_{k}\langle q_{k}\rangle\langle
F_{k}\rangle\:.
\end{equation}
 The
auto-correlation and the cross-correlation functions for the
currents can be expressed as

\begin{eqnarray}\label{cross-corr-curr}
\langle i_{k}(t)i_{j}(t+\tau)\rangle =\sum_{n,m}\langle
q_{kn}q_{jm}
f(t-t_{kn})f(t-t_{jm}+\tau)\rangle\nonumber\\
=\int\int dt_{k}dt_{j} \langle q_{k} q_{j}\rangle
f(t-t_{k})f(t-t_{j}+\tau)\langle
F_{k}(t_{k})F_{j}(t_{j})\rangle,\nonumber\\
\end{eqnarray}
respectively for $k=j$ where $\langle
F_{k}(t_{k})F_{k}(t'_{k})\rangle$ is the auto-correlation function
of the photon flux at detector $k$, and for $k\neq j$ where
$\langle F_{k}(t_{k})F_{j}(t_{j})\rangle$ is the cross-correlation
between the fluxes incident at the two different detectors.

\subsection{Analog calibration using SPDC}

The functions $\langle F_{k}(t_{k})F_{k}(t'_{k})\rangle$ and
$\langle F_{k}(t_{k})F_{j}(t_{j})\rangle$ are evaluated in
\cite{systematic} for SPDC. As mentioned above, the resolving time
of a real analog detector is finite, and in general can be
considered much larger than the SPDC coherence time. Thus any
fluctuations in the intensity of light are integrated over
$\tau_{p}$ during the detection process. In the limit of
$\tau_{p}\gg\tau_{coh}$, for $k=1$ and $j=2$ we have

\begin{eqnarray}\label{self-corr-curr2}
\langle i_{1}(t) i_{1}(t+\tau)\rangle=\langle
i_{1}\rangle^{2}+\eta_{1}\langle q_{1}^{2}\rangle
\mathcal{F}(\tau) \cdot\left[\langle
F_{1}\rangle+\eta_{1}\Im\right],
\end{eqnarray}
\begin{eqnarray}\label{cross- corr-curr2}
\langle i_{1}(t)i_{2}(t+\tau)\rangle=\langle i_{1}\rangle\langle
i_{2}\rangle+\eta_{1}\eta_{2}\langle q_{1}\rangle\langle
q_{2}\rangle \mathcal{F}(\tau) \cdot\left[\langle
F_{1}\rangle+\Im\right]\nonumber\\
\end{eqnarray}
where we have introduced the convolution of the response function
of detectors $\mathcal{F}(\tau)\equiv\int dt f(t)f(t+\tau)$. The
term $\Im$ depends on the second power of the parametric gain $V$,
i.e. the mean number of photons per mode of the radiation. For our
purposes we can estimate approximatively $\Im\simeq V\langle
F\rangle$ \cite{systematic}. The  Eqs (\ref{self-corr-curr2}) and
(\ref{cross- corr-curr2}) are the fundamental tools for studying
the problem of absolute calibration of analog detectors and thus
we are going to discuss them in detail. Despite they seem to be
quite symmetric, we observe an important difference. The term
proportional to $\langle F_{1}\rangle$ in the autocorrelation
function is the shot noise contribution and for this reason the
quantum efficiency $\eta_{1}$ enters linearly, while in the
current cross-correlation function the corresponding term is due
to the high quantum correlation between the signal and idler beams
of PDC and the quantum efficiency appears quadratically. Its
presence is the key for absolute calibration. The term $\Im$ both
for auto- and cross-correlation functions, can be neglected as
long as $V\ll 1$, i.e., the mean number of photons per mode is
much smaller than one. However, if the duration of the
photocurrent pulse is much longer than the coherence time, this
assumption does not prevent photodetection being in a strongly
analog regime, because a lot of photons can be absorbed during the
pulse duration generating overlapping of pulses [see Eq
(\ref{curr})]. The term proportional to $\langle
i_{}\rangle^{2}\propto\langle F_{}\rangle^{2}$ is due to the
presence of more than one photon within a time interval $\tau_{p}$
and for that reason is more delicate. It can be neglected in Eqs.
(\ref{self-corr-curr2}) and (\ref{cross- corr-curr2}) only if
$\langle F_{}\rangle\ll\mathcal{F}(\tau)$. Since the pulse $f(t)$
has a height around $1/\tau_{p}$,
$max\left[\mathcal{F}(\tau)\right]=\mathcal{F}(0)\sim 1/\tau_{p}$.
So the condition becomes $\langle F_{}\rangle\tau_{p}\ll 1$, i.e.,
the number of incident photons during the resolving time of the
detector should be much less than one, i.e.  one should work in a
non-overlapping regime.

Therefore, it is convenient to distinguish among three different
regimes:

\textbf{(I)} $\langle F_{}\rangle\tau_{p}\ll 1$ (i.e. photocurrent
pulses do not overlap on the average). For a detector with a time
constant $\tau_p=10$ ns, the corresponding photon flux must be
below $10^{8}$ photons/s, which, for a wavelength of $500$ nm,
means a power of tens of pW. From (\ref{self-corr-curr2}) and
(\ref{cross- corr-curr2}), neglecting $\Im$ and the terms in
$\langle i\rangle^{2}$ one obtains a formula for the estimation of
quantum efficiency like the one reported in \cite{klysh}. However,
this regime does not appear interesting because of the necessity
to work only at very low intensity, where no overlapping between
pulses happens. In principle, one could distinguish between
different pulses of the current and work in the counting mode
provided the amplitude $q_{n}$ of each pulse is large enough to be
detectable.

\textbf{(II)} $\langle F_{}\rangle\tau_{p}\gtrsim 1$ but still
$V\ll 1$ (i.e., photocurrent pulses overlap but the parametric
gain and photon flux are still quite low). By considering the same
parameters as used in case I, coherence time $\tau_{coh}$ of the
order of $100$ fs, and the requirement that $V\leq 0.001$, this
means a photon flux up to $10^{10}$ photons/s, i.e. a power of few
nW. Here, only the term $\Im$ can be neglected in Eqs.
(\ref{self-corr-curr2}) and (\ref{cross- corr-curr2}). Defining
the current functions as $ \delta i_{k}(t)\equiv i_{k}(t)-\langle
i_{k}\rangle$, the analog quantum efficiency can be estimated as:
\begin{equation}\label{eta-klyshko-MI}
\eta_{2}\langle q_{2}\rangle=\frac{\langle
q_{1}^{2}\rangle}{\langle q_{1}\rangle^{2}}\langle q_{1}\rangle
  \frac{\langle \delta i_{1}(t)\delta i_{2}(t+\tau)\rangle}{\langle
\delta i_{1}(t) \delta
  i_{1}(t+\tau)\rangle}.
\end{equation}
Unfortunately it is not satisfying from the metrology point of
view because of the presence of an unknown parameter $M=\langle
q_{1}^{2}\rangle/\langle q_{1}\rangle^{2}$ related to the
statistics of charge fluctuations and that has to be estimated
independently in some other way. We suggest to avoid the problem
by integrating Eq. (\ref{cross- corr-curr2}) over time $\tau$. It
correspond to evaluate the power spectrum of the fluctuations at
frequencies around zero, namely much smaller than $1/\tau_{p}$. We
would like to stress that in this case, the assumption
$f_{1}(t)=f_{2}(t)=f(t)$ is not necessary. Since $\int
d\tau\mathcal{F}(\tau)=1$, integrating Eq. (\ref{cross-
corr-curr2}) in $\tau$ and dividing by the current [see Eq
(\ref{curr-time})], we obtain
\begin{equation}\label{eta-klyshko-integ-MI}
\eta_{2}\langle q_{2}\rangle=\frac{\int d\tau\langle \delta
i_{1}(t)\delta i_{2}(t+\tau)\rangle}{\langle i_{1}\rangle}\:.
\end{equation}
Here $\eta_{2}\langle q_{2}\rangle$ is the electron charge
produced per single incident photon, or, according to formula
(\ref{curr-time}), the ratio between the current and the photon
flux.

Eq. (\ref{eta-klyshko-integ-MI}) represents one of the main
results of the paper, since it shows that the absolute calibration
of analog detectors by using SPDC is indeed possible.

\textbf{(III)} $V\gtrsim 1$ (fluxes larger than $10^{10}$
photons/s). In this regime each term of (\ref{self-corr-curr2})
and (\ref{cross- corr-curr2}) is important and no simple way can
be found for the absolute calibration with SPDC. It can be shown
that in this case one should be able to collect exactly the same
number of correlated modes by $D1$ and $D2$ \cite{systematic}.
Since SPDC takes place with a very large spectral and spatial
bandwidth, it would require accurate and well balanced spatial and
frequency selection. This could originate systematic errors which
are difficult to be evaluated.

\subsection{Analog calibration using stimulated PDC}

We overcome the problem of calibration for photon fluxes larger
than $10^{10}$, highlighted in point (III), exploring the
possibility of using a parametric amplifier configuration, where a
seed coherent beam with power $\phi$ is injected along direction 2
stimulating the bright emission of the two correlated beams. In
this case the photon fluxes can be increased by varying the power
$\phi$ of the coherent seed beam, without increasing $V$ (we can
lead back to the condition $V\ll1$). Introducing the proper
expression for the fluxes in the stimulated emission in
(\ref{curr-time}), under the condition that the spontaneous
emission is negligible we have
\begin{eqnarray}\label{stimul-curr}
\langle i_{1}\rangle_{\mathrm{s}} &=&\eta_{1}\langle q_{1}\rangle
V\phi
\nonumber\\
\langle i_{2}\rangle_{\mathrm{s}} &=& \eta_{2}\langle
q_{2}\rangle(1+V)\phi
\end{eqnarray}
where the subscript $\mathrm{s}$ reminds that here the currents
are calculated for the stimulated PDC. Using Eq.
(\ref{cross-corr-curr}) the correlation functions of the current
fluctuations can be expressed in a simple form, under the
assumption of $V\ll1$:

\begin{eqnarray}\label{stimul-corr}
\langle \delta i_{1}(t) \delta
i_{1}(t+\tau)\rangle_{\mathrm{s}}&=&\eta_{1}\langle
q_{1}^{2}\rangle
\mathcal{F}(\tau) V\phi\nonumber\\
\langle \delta i_{2}(t) \delta
i_{2}(t+\tau)\rangle_{\mathrm{s}}&=& \eta_{2}\langle
q_{2}^{2}\rangle \mathcal{F}(\tau) (1+V)\phi\\
\langle \delta i_{1}(t)\delta i_{2}(t+\tau)\rangle_{\mathrm{s}}
&=& 2\eta_{1}\eta_{2}\langle q_{1}\rangle\langle q_{2}\rangle
\mathcal{F}(\tau) V\phi\nonumber
\end{eqnarray}

The quantum efficiency can be evaluated as
\begin{equation}\label{eta-stim-tau}
\eta_{2}\langle q_{2}\rangle=\frac{1}{2}\frac{\langle
q_{1}^{2}\rangle}{\langle q_{1}\rangle^{2}}\langle q_{1}\rangle
  \frac{\langle \delta i_{1}(t)\delta i_{2}(t+\tau)\rangle_{\mathrm{s}}}{\langle
\delta i_{1}(t) \delta
  i_{1}(t+\tau)\rangle_{\mathrm{s}}}.
\end{equation}

Alternatively, integrating in $\tau$ the expression for the
cross-correlation we have
\begin{equation}\label{eta-stim}
\eta_{2}\langle q_{2}\rangle=\frac{1}{2}\frac{\int d\tau\langle
\delta i_{1}(t)\delta i_{2}(t+\tau)\rangle_{\mathrm{s}}}{\langle
i_{1}\rangle_{\mathrm{s}}}\:,
\end{equation}
in which the avalanche gain factor disappears. There are two main
advantages of using the stimulated PDC. First of all we are not
up-limited in the photon fluxes that we can use. Considering
detector without internal gain where the Johnson noise in the
amplification process is predominant, it is important to obtain a
good signal to noise ratio by increasing the signal. We stress
that in our case the signal is of the order of the shot noise! A
second advantage lies in the very narrow bandwidth, both in space
and frequency of the stimulated emission, that is the one of the
coherent seed beam. This makes easier to collect the correlated
beams resulting in two bright spots in the detection plane.

\section{Preliminary uncertainty analysis}

The calculation of the expected uncertainties in the evaluation of
the correlations functions depends on the measurement technique.
Here we consider a measurement process in which the two current
outputs of $D1$ and $D2$ are combined by an analog multiplier and
the result is integrated for a time $T$ much larger than the
response time $\tau_{p}$. The general calculation of the variance
of this quantity would require the knowledge of the fourth-order
quantum correlation functions of the photon fluxes which are
rather complex. Therefore we used the approximation of small
fluctuations, i.e. $\delta i(t)\ll \langle i\rangle$, which
perfectly fits the regime of strong overlapping discussed in the
previous section. In this case one obtains, both for the auto- and
cross-correlation, a maximal uncertainty

\begin{eqnarray}\label{delta-corr}
\underline{\Delta}_{\mathrm{corr}}\approx \eta^{2}\langle
q\rangle^{2} \langle F\rangle^{3/2}T^{-1/2}
\end{eqnarray}

Considering equation (\ref{eta-klyshko-MI}) for a detector without
avalanche gain, following the uncertainty propagation rule, we
obtain the relative uncertainty on the quantum efficiency as

\begin{eqnarray}\label{delta-eta}
\frac{\underline{\Delta}_{\eta}}{\eta}\approx \langle
N_{\tau_{p}}\rangle^{1/2}\left(\frac{\tau_{p}}{T}\right)^{1/2}
\end{eqnarray}
where $N_{\tau_{p}}$ is the number of photons detected during  the
detector response time. The uncertainty scales with the square
root of the measurement time $T$. For example, if $T=1$ s, it is
lower than one part in $10^{-3}$. Since, as we are going to
discuss, this term is not the dominant one this estimate suffices
for our purposes.

The noise contributions from the detector (dark current, noise of
the transimpedance amplifier) and background light are supposed to
be statistically independent of each other and of the
photocurrents produced by SPDC light in the two detectors. Thus,
they do not give any contribution to the cross correlation
function of the two current fluctuations in the numerator of
(\ref{eta-klyshko-MI}). In the denominator of the same equation,
containing the autocorrelation of the current fluctuations, their
effect should be measured (for example, rotating by $90^{o}$ the
polarization of the laser pump, to eliminate the SPDC signal), and
subtracted. The statistical uncertainty of this measurement can be
kept so small that it does not increase significantly the total
relative uncertainty given above.

The main systematic contribution is due to the non linear crystal
optical losses. The uncertainty in its measurement could be
reduced to few parts in $10^{3}$ \cite{DMBDMS,BCDNR}. Thus, it is
the dominant contribution to the uncertainty budget, largely
exceeding the statistical one.

\section{Conclusion}

Motivated by the necessity of a general absolute calibration
scheme for detectors for various applications, we have performed a
systematic study on the possibility of applying PDC calibration
methods to the analog regime. Our results show that the
measurement of correlations between detector output currents can
indeed be used to extend the absolute calibration method to the
analog regime. In particular, it is shown that integration of the
photocurrent correlation functions in time allows one to avoid the
measurement of the photocurrent pulse shape and to eliminate the
necessity to know the statistics of the internal gain of the
detector. Also, our analysis shows that is possible to go beyond
the regime of non-overlapping photocurrent pulses, which was used
in earlier works, and to move on to higher intensities.
Furthermore, the stimulated operating mode of PDC has been
considered with promising perspectives.

A preliminary theoretical uncertainty analysis gives a relative
value around $10^{-3}$, mainly due to the measurement uncertainty
of the optical loss in the crystal, which is interesting from a
metrology view-point.





%

\end{document}